\documentclass[10pts,showpacs,preprint,aps]{revtex4}
\usepackage{graphicx}
\usepackage{dcolumn}
\usepackage{bm}
\usepackage{amssymb}
\usepackage{amsmath}
\begin{document}
\setcounter{page}{1}
\vskip 2cm
\title
{The Nambu-Goldstone theorem in non-relativistic systems}
\author
{Ivan Arraut}
\affiliation{Department of Physics, Faculty of Science, Tokyo University of Science,
1-3, Kagurazaka, Shinjuku-ku, Tokyo 162-8601, Japan}

\begin{abstract}
In non-relativistic systems, when there is spontaneous symmetry breaking, the number of Nambu-Goldstone bosons ($n_{NG}$) are not necessarily equal to the number of broken generators ($n_{BG}$). Here we use the method of operators for analyzing the necessary conditions in order to obtain the correct dispersion relation for the Nambu-Goldstone bosons.                                 
\end{abstract}
\maketitle 
\section{Introduction}
The Nambu-Goldstone theorem suggests that when the symmetry is spontaneously broken, the number of Nambu-Goldstone bosons is equivalent to the number of broken generators \cite{Goldstone, Godstone2, Nambu, Goldstone3}. This is the case in relativistic systems. In non-relativistic systems, this equality is not necessarily satisfied. This issue has been analyzed before in the seminal work of Nielsen and Chadha \cite{NC}, and subsequently in a remarkable and simple way by Watanabe, Murayama and Brauner \cite{Mura}. The idea is that for the case where the relativistic invariance is violated, some broken generators of the theory become canonically conjugate to each other, such that the corresponding Nambu-Goldstone bosons represent a single degree of freedom. Another consequence of the absence of relativistic invariance is the fact that the Nambu-Goldstone bosons corresponding to the canonically conjugate pairs, have in general a quadratic dispersion relation $\omega\backsim \vec{k}^2$ instead of the expected linear dispersion relation corresponding to mass-less particles $\omega\backsim \vec{k}$. This issue was analyzed in \cite{L} by using the effective field theory approach and in \cite{Nico} by using other methods. In \cite{B2} a more elaborated proof of the changes in the dispersion relations for the Goldstone bosons by analyzing linear sigma models with chemical potentials at tree level was done. The same problem analyzed from the point of view of Kaon condensation in QCD was developed in \cite{Kaon}.  
In this paper, we analyze the condition for spontaneous symmetry breaking $<0\vert[\phi_i(x),Q_a(y)]\vert0>\neq0$, but expressed in the form $<0\vert[\phi_i(x),[Q_a(y),Q_b(z)]]\vert0>\neq0$. This condition together with the additional condition over the commutation among the broken generators $<0\vert[Q_a(z),Q_b(y)]\vert0>$, are enough for finding the appropriate dispersion relations for the Nambu-Goldstone bosons and the appropriate number. Although this issue has already been analyzed partially in \cite{NC}, here we develop an alternative derivation by writing the necessary conditions in order to get a quadratic dispersion relations. The final condition analyzed in this paper is equivalent to say that the full analysis developed by Nielsen and Chadha in \cite{NC} can be condensed in the result $<0\vert[\phi_i(x),[Q_a(y),Q_b(z)]]\vert0>\neq0$. The paper is organized as follows: In Sec. (\ref{Se1}), we make a review of the standard Nambu-Goldstone theorem for relativistic systems. In Sec. (\ref{eq:sec1}), we analyze the dispersion relations for the Nambu-Goldstone bosons for two cases, namely, {\bf i).} The case where the broken generators do not form conjugate pairs. {\bf ii).} The case where the broken generators form conjugate pairs. Finally, in Sec. (\ref{Sec6}), we conclude. 

\section{The Nambu-Goldstone theorem: Standard case}   \label{Se1}

If we define a potential depending on some field $\phi^i$ as $V(\phi^i)$, then it can be expanded in the following way

\begin{equation}   \label{eq:invariant}
V(\phi)=V(\phi_0)+\frac{1}{2}(\phi-\phi_0)^a(\phi-\phi_0)^b\left(\frac{\partial^2}{\partial\phi^a\partial\phi^b}V\right)_{\phi_0}+....,
\end{equation}  
where $\partial V/\partial\phi^a=0$, since we are expanding around an extremal point. The mass matrix is symmetric and given by the coefficient

\begin{equation}
\left(\frac{\partial^2}{\partial\phi^a\partial\phi^b}V\right)_{\phi_0}=m^2_{ab}\geq0.
\end{equation}
This previous condition is due to the fact that $\phi_0$ represents a minimum. At this point we will assume that the full action given by

\begin{equation}
\pounds=(kinetic\;\; terms)-V(\phi),
\end{equation} 
is invariant under the the action of the group $G=O(N)$. In addition, we assume that the selected vacuum state is invariant under the action of a subgroup of $G$, given by $H=O(N-1)$. The vacuum state is not invariant under the action of the full group $G$. In summary

\begin{equation}
G:\;\;\phi_0 ^{a'} =U(g)\phi_0 ^a\neq\phi_0^a,
\end{equation} 

\begin{equation}
H:\;\;\phi_0^{a'}=U(h)\phi_0 ^a=\phi_0^a,
\end{equation}
where $U(g)$ and $U(h)$ denote the representations of the groups $G$ and $H$ respectively. However, the potential $V(\phi)$ is still invariant under the action of the full group $G$. The action of this group on the potential expansion (\ref{eq:invariant}) gives the result

\begin{equation}   \label{eq:Globalsymmetry}
T^a(\phi)\frac{\partial}{\partial\phi^a}V(\phi)=0,
\end{equation}
where $T^a(\phi)$ are the generators of the group of transformations. In the $U(g)$ representation for example, it would take the form

\begin{equation}
U(g)=e^{T^a\alpha}\approx \hat{I}+\alpha T^a\to U(h)\phi_0^a=\phi_0^a+\alpha T^a(\phi_0),
\end{equation}
where $T^a(\phi)$ denotes the action of the operator $T^a$ on the function state $\phi$. If we introduce the result (\ref{eq:Globalsymmetry}) inside the expansion (\ref{eq:invariant}), then we get the condition

\begin{equation}   \label{eq:Globalsymmetry2}
T^a(\phi_0)T^b(\phi_0)\frac{\partial^2}{\partial\phi^a\partial\phi^b}V(\phi)=T^a(\phi_0)T^b(\phi_0)m^2_{ab}=0.
\end{equation}
Note that if $T^a(\phi_0)=0$, namely, when the vacuum state selected is invariant under the action of the group, then the corresponding mass component $m_{ab}$ is not necessarily zero. On the other hand, when the symmetry generator is broken, namely, when the group element belonging to $G$ does not leave the vacuum invariant, then $T^a(\phi_0)\neq0$ and then the mass components related to this condition are necessarily zero ($m_{ab}=0$). Then the number of broken generators are clearly related to the existence of the Nambu-Goldstone bosons. In summary, the number of fields whose mass is not required to be zero, is determined by the dimension of the subgroup $H$ under which the vacuum is invariant. On the other hand, the number of Nambu-Goldstone bosons (mass-less particles) is determined by the dimension of the coset $G/H$. The results of this section can be found in any textbook, for a reference, please see \cite{Field1}. 

\subsection{Charge conservation}

At the quantum level, the Goldstone theorem suggests that if there is a field operator $\phi(x)$ with non-vanishing vacuum expectation value $<0\vert\phi(x)\vert0>$ and in addition, the vacuum expectation value is not a singlet under the transformations of some representation of a symmetry group, then some mass-less particles will appear in the spectrum states. It is well known that up to a total derivative term in the action (Lagrangian $\pounds$), the conserved charge is given by

\begin{equation}
j^a_\mu(x)=\frac{\partial\pounds}{\partial(\partial^\mu\phi)}\frac{\delta\phi(x)}{\delta\alpha^a},
\end{equation}
where $\delta\phi(x)/\delta\alpha$ corresponds to the field variations under symmetry transformations of the Lagrangian. The previously defined current is divergence-less and the corresponding charges are given by

\begin{equation}   \label{eq:Chargedef}
Q^a(x)=\int d^3xj^a_0(x).
\end{equation}  
In the standard cases, these charges are conserved, $dQ^a/dt=0$, and they have a well defined commutation relations defined in agreement with the Lie algebra

\begin{equation}   \label{Liealgebralalala}
[Q^a,Q^b]=C^{abc}Q^c,
\end{equation}
where $C^{abc}$ are the structure constants of the Lie algebra. We can define an unitary operator with the charge being the generator of the group transformations

\begin{equation}
U=e^{iQ^a\alpha^a}.
\end{equation}
If the vacuum is non-degenerate, then the previously defined charge annihilates the vacuum, namely, $U\vert0>=\vert0>$, or equivalently

\begin{equation}
Q^a\vert0>=0.
\end{equation}    
When the vacuum is degenerate, then these previous conditions are not satisfied and in general the charges cannot annihilate the vacuum

\begin{equation}
U\neq e^{iQ^a\alpha^a}, \;\;\;\;\;Q^a\vert0>\neq0. 
\end{equation}
If the operator $\phi(x)$ is not a singlet under the action of the group, then its commutation with the charge $Q^a$ is non-zero and defined by

\begin{equation}   \label{eq:brokencondition}
[Q^a,\phi'(x)]=\phi(x).
\end{equation}    
The vacuum expectation value of this operator is given by

\begin{equation}   \label{eq:brokencondition2}
<0\vert Q^a\phi'(x)-\phi'(x)Q^a\vert0>\neq0.
\end{equation}
If we introduce the definition of charge given in eq. (\ref{eq:Chargedef}), then we get

\begin{equation}    \label{eq:sum}
\sum_n\int d^3y\left[<0\vert j^a_0(y)\vert n><n\vert\phi'(x)\vert0>-<0\vert\phi'(x)\vert n><n\vert j^a_0(y)\vert0>\right]_{x^0=y^0}\neq0,
\end{equation}
where the equal-time condition has been imposed. If we impose the translational invariance condition

\begin{equation}   \label{expo}
 j^a_0(y)=e^{-ipy} j^a_0(0)e^{ipy},
\end{equation}
and then we replace it inside eq. (\ref{eq:sum}), we obtain

\begin{equation}   \label{eq:sum2}
(2\pi)^3\sum_n\delta^{(3)}(\vec{p}_n)\left[<0\vert j^a_0(0)\vert n><n\vert\phi'(x)\vert0>e^{iM_ny^0}-<0\vert\phi'(x)\vert n><n\vert j^a_0(0)\vert0>e^{-iM_ny^0}\right]_{x^0=y^0}\neq0.
\end{equation}
Note that the delta functions come from the spatial integrations of the exponential defined in eq. (\ref{expo}). The previous condition must be different from zero in agreement with the spontaneous symmetry breaking argument. Note that $p_n^0=M_n$ and the current conservation guarantees that the previous expression is independent of $y_0$. This is the case if the condition $M_n=0$ is satisfied and this proves the Goldstone theorem. Note that the previous expressions implies that $M_n$ and $\vec{p}_n$ go to zero simultaneously. For a mass-less particle this implies a dispersion relation of the form $\omega\backsim \vec{k}$, this dispersion relation appears naturally from the condition (\ref{eq:sum2}). The proof elaborated in the previous way can be found in \cite{Field2}. From the definition of delta Dirac, given by the integral

\begin{equation}
\int d^3ye^{i\vec{p}\cdot\vec{y}}=(2\pi)^3\delta^{(3)}(\vec{p}),
\end{equation} 
it is possible to understand that at the lowest order in the expansion (long wavelength regime), the momentum dispersion is linear. This fact can be observed from the expansion of the exponential term in the limit $\vec{p}\to0$ defined as

\begin{equation}
e^{i\vec{p}\cdot\vec{y}}=\sum_{r=0}^\infty\frac{\left(i\vec{p}\cdot\vec{y}\right)^r}{r!},
\end{equation} 
which at the lowest order (long wavelengths) has a linear dependence on $\vec{p}$. This explains why for relativistic systems, or more generally, for systems where the broken generators do not form conjugate pairs, the Nambu-Goldstone bosons have the expected dispersion relation $\omega\backsim\vec{k}$ or equivalently $E\backsim \vec{p}$. Later we will demonstrate that this is not the case if the broken generators form canonical conjugate pairs. The spontaneous symmetry breaking condition (\ref{eq:brokencondition2}), can be expressed in a more explicit form as

\begin{equation}   \label{impo}
<0\vert[\phi_i,[Q_\alpha,Q_\beta]]\vert0>\neq0.
\end{equation}
Here we take into account the Lie algebra structure defined in eq. (\ref{Liealgebralalala}) for the broken generators. In such a case, there is no difference between the condition (\ref{impo}) and the one showed previously in eq. (\ref{eq:brokencondition2}), as far as the structure constants of the Lie algebra are considered fixed.     

\section{Spontaneous symmetry breaking in non-relativistic systems}   \label{eq:sec1} 

From the results of the previous section, spontaneous symmetry breaking at the quantum level is defined by the relation between the order parameter of the theory and the broken generators. This relation is summarized in the matrix

\begin{equation}   \label{eq:eq1la}  
M_{ab}=<0\vert [Q_a, \Phi_b(0)]\vert0>,
\end{equation}
which has to be a non-singular matrix if we want to satisfy the condition of spontaneous symmetry breaking \cite{Mura}. The set of generators, satisfying the previous condition are recognized as the broken generators of the theory. Given a number of broken generators $n_{BG}$, the Nambu-Goldstone theorem developed in the previous section suggests that $n_{BG}=n_{NG}$, where $n_{NG}$ corresponds to the number of Nambu-Goldstone bosons. It has been demonstrated that for the case where the relativistic invariance is absent, the appropriate relation between the number of Nambu-Goldstone bosons and the number of broken generators is \cite{Mura, Murala}  

\begin{equation}   \label{eq:eq1la2}
n_{BG}-n_{NG}=\frac{1}{2}Rank(\rho), 
\end{equation}
where the matrix $\rho$ is defined by

\begin{equation}   \label{eq:eq1la3}
\rho_{ab}=-iLim_{V\to\infty , B_i\to0}\frac{1}{V}<0\vert [Q_a, Q_b]\vert0>.
\end{equation}
Here $V$ corresponds to the spatial volume and $B_i$ are the external fields which help to the order parameters to get a preferred direction. This formula has been demonstrated in \cite{Mura, Murala}, inspired in the previous results obtained by Nielsen and Chadha \cite{NC}.  

\subsection{Second constraint}

By using similar arguments with respect to Sec. (\ref{Se1}), we can start from the result

\begin{equation}   \label{eqconstr}
<0\vert Q_\alpha Q_\beta-Q_\beta Q_\alpha\vert0>\neq0,
\end{equation}               
and then we define the charge density with eq. (\ref{eq:Chargedef}) and after introducing a complete set of intermediate states, this previous expression becomes

\begin{equation}
\sum_n\int d^3yd^3x\left(<0\vert j_{0\alpha}(y)\vert n><n\vert j_{0\beta}(x)\vert0>-<0\vert j_{0\beta}(x)\vert n><n\vert j_{0\alpha}(y)\vert0>\right)\neq0.
\end{equation}					
Note that now we have two spatial integrals instead of a single one. If we keep the condition (\ref{expo}) for both currents, then the previous equation becomes

\begin{eqnarray}   \label{the ;ast one2}
\sum_n\int d^3yd^3x(<0\vert j_{0\alpha}(0)\vert n><n\vert j_{0\beta}(0)\vert0>e^{i(M_ny^0-\tilde{M}_nx^0)}e^{-i(\vec{p}_n\cdot\vec{y}-\vec{\tilde{p}}_n\cdot\vec{x})}\nonumber\\
-<0\vert j_{0\beta}(0)\vert n><n\vert j_{0\alpha}(0)\vert0>e^{-i(M_ny^0-\tilde{M}_nx^0)}e^{i(\vec{p}_n\cdot\vec{y}-\vec{\tilde{p}}_n\cdot\vec{x})})\neq0.
\end{eqnarray}
If we impose the condition of equal time commutation relations, then $y^0=x^0$. In addition, if we make the spatial integrals, we obtain the result

\begin{eqnarray}   \label{the ;ast one}
(2\pi)^6\sum_n\delta^{(3)}(\vec{p}_n)\delta^{(3)}(-\vec{\tilde{p}}_n)(<0\vert j_{0\alpha}(0)\vert n><n\vert j_{0\beta}(0)\vert0>e^{iy^0(M_n-\tilde{M}_n)}\nonumber\\
-<0\vert j_{0\beta}(0)\vert n><n\vert j_{0\alpha}(0)\vert0>e^{-iy^0(M_n-\tilde{M}_n)})\neq0.
\end{eqnarray}
Note that in any case the delta-Dirac function is symmetric in its argument. The previous result is non-trivial, only if the condition

\begin{equation}   \label{disp}
M_n=\tilde{M}_n,
\end{equation}
is satisfied. Note that here we use different symbols for the quantities corresponding to different generators. At the end the condition (\ref{eqconstr}), suggests that broken generators forming conjugate pairs, correspond to the same degree of freedom. At this point it is not possible to say anything about the dispersion relation of the Nambu-Goldstone boson. However, for sure we can say that $\vec{p}_n=\vec{\tilde{p}}_n$ in eq. (\ref{the ;ast one}), which is consistent with the result (\ref{disp}). 

\subsection{The combined conditions}

If we combine the conditions (\ref{eq:brokencondition2}) and (\ref{eqconstr}), then we will be able to reproduce the appropriate dispersion relation. Here we develop the spontaneous symmetry breaking condition, but expressed in the form (\ref{impo}). By developing such expression and using then the constraints obtained from eq. (\ref{disp}), we must be able to find the appropriate dispersion relation for the Nambu-Goldstone bosons. If the results exposed in \cite{Mura} are consistent, then in the absence of the condition (\ref{eqconstr}), the dispersion relation is linear. Here we will develop the dispersion relations for different cases.  
 
\subsubsection{i). All the broken generators completely independent and not forming canonical conjugate pairs}   \label{thissubsectionla}    

For this case, the condition (\ref{eqconstr}) becomes trivial and the $\rho$ matrix defined in eq. (\ref{eq:eq1la3}) vanishes. Note that if eq. (\ref{the ;ast one}) vanishes as a consequence of the triviality imposed here over eq. (\ref{eqconstr}), then the condition (\ref{disp}) is not satisfied. In fact, it is clear that in order to get a trivial solution for eq. (\ref{the ;ast one}), we need to satisfy

\begin{equation}   \label{keycond}
M_n\neq \tilde{M}_n. 
\end{equation}  
Under such condition, eq. (\ref{the ;ast one}) would vanish after average over a large period of time. However, the spontaneous symmetry breaking condition, must be satisfied anyway. From eq. (\ref{impo}) and ignoring for the moment spatial integrations, we obtain

\begin{eqnarray}
<0\vert\phi_i(z)\vert n><n\vert\ j_{0\alpha}(x)\vert n'><n'\vert j_{0\beta}(y)\vert0>-<0\vert\phi_i(z)\vert n><n\vert\ j_{0\beta}(x)\vert n'>\times\nonumber\\
<n'\vert j_{0\alpha}(y)\vert0>-<0\vert\ j_{0\alpha}(x)\vert n'><n'\vert j_{0\beta}(y)\vert n><n\vert\phi_i(z)\vert 0>+\nonumber\\
<0\vert\ j_{0\beta}(x)\vert n'><n'\vert j_{0\alpha}(y)\vert n><n\vert\phi_i(z)\vert 0>\neq 0.
\end{eqnarray} 
Here the sum over $n$ and $n'$ is understood and we omit the spatial integration symbols over the variables $x$ and $y$ but it must be understood that in fact the integrations over these variables must be done. If again we assume symmetry under space-time translations by using the result (\ref{expo}), then we obtain

\begin{eqnarray}   \label{eq:strange}
<0\vert\phi_i(z)\vert n><n\vert\ j_{0\alpha}(0)\vert n'><n'\vert j_{0\beta}(0)\vert0>e^{-i(M_n-M_{n'})x^0}e^{-i\tilde{M}_{n'}y^0}e^{i(\vec{p_n}-\vec{p_{n'}})\cdot\vec{x}}e^{i(\tilde{\vec{p}}_{n'})\cdot\vec{y}}\nonumber\\
-<0\vert\phi_i(z)\vert n><n\vert\ j_{0\beta}(0)\vert n'><n'\vert j_{0\alpha}(0)\vert0>e^{-i(\tilde{M}_n-\tilde{M}_{n'})y^0}e^{-iM_{n'}x^0}e^{i(\tilde{\vec{p_n}}-\tilde{\vec{p_{n'}}})\cdot\vec{y}}e^{i(\vec{p}_{n'})\cdot\vec{x}}\nonumber\\
-<0\vert\ j_{0\alpha}(0)\vert n'><n'\vert j_{0\beta}(0)\vert n><n\vert\phi_i(z)\vert 0>e^{-i(\tilde{M}_{n'}-\tilde{M}_{n})y^0}e^{iM_{n'}x^0}e^{i(\tilde{\vec{p_{n'}}}-\tilde{\vec{p_{n}}})\cdot\vec{y}}e^{-i(\vec{p}_{n'})\cdot\vec{x}}
\nonumber\\
+<0\vert\ j_{0\beta}(0)\vert n'><n'\vert j_{0\alpha}(0)\vert n><n\vert\phi_i(z)\vert 0>e^{-i(M_{n'}-M_n)x^0}e^{i\tilde{M}_{n'}y^0}e^{i(\vec{p_{n'}}-\vec{p_{n}})\cdot\vec{x}}e^{-i(\tilde{\vec{p}}_{n'})\cdot\vec{y}}\nonumber\\
\neq 0.
\end{eqnarray}    
In this case, if we want to keep the non-trivial result, we have to guarantee invariance under time-translations. This is equivalent to the condition $M_n=M_{n'}$, corresponding to the degree of freedom of the first broken generators, as well as $\tilde{M}_n=\tilde{M}_{n'}$, corresponding to the degree of freedom for the second broken generator. Here in addition, the condition (\ref{keycond}) is satisfied. Note that all the spacetime dependence appears inside the phases (exponential functions). Then for a system invariant under time-reversal, we can simplify the expression (\ref{eq:strange}) as follows

\begin{eqnarray}   \label{eq:strange77lala}
\sum_{n,n'}C_{n,n'}<0\vert\phi_i(z)\vert n><n'\vert j_{0\beta}(0)\vert0>\left(e^{-i\tilde{M}_{n'}y^0}e^{i(\tilde{\vec{p}}_{n'})\cdot\vec{y}}+e^{i\tilde{M}_{n'}y^0_R}e^{-i(\tilde{\vec{p}}_{n'})\cdot\vec{y}}\right)\nonumber\\
-D_{n,n'}<0\vert\phi_i(z)\vert n><n'\vert j_{0\alpha}(0)\vert0>\left(e^{-iM_{n'}x^0}e^{i(\vec{p}_{n'})\cdot\vec{x}}+e^{iM_{n'}x^0_R}e^{-i(\vec{p}_{n'})\cdot\vec{x}}\right)\neq 0,
\end{eqnarray}
where we have defined $C_{n,n'}=<n\vert\ j_{0\alpha}(0)\vert n'>$ and $D_{n,n'}=<n'\vert j_{0\alpha}(0)\vert n>$ with the corresponding adjoints. Here the invariance of the system under time-reversal, guarantees that $y^0_R=y^0$ and $x^0_R=x_0$, where we identify $y^0_R$ and $x^0_R$ as the time-reversal version of time. Due to the inequality defined in eq. (\ref{keycond}), it is not possible to factorize more the previous expression. By taking into account the previous considerations, it is clear that each broken generator identifies different degrees of freedom. The sum of the exponential terms in eq. (\ref{eq:strange77lala}) at the lowest order, gives a linear dependence in both, position and momentum, defining then a linear dispersion relation $\omega\backsim \vec{k}$. Note that the linear dispersion relation comes from the fact that the system is invariant under time-reversal. On the other hand, the fact that the number of Nambu-Goldstones is equivalent to the number of broken generators, comes from the result (\ref{keycond}).     

\subsubsection{ii). All the broken generators forming conjugate pairs}

In this case where the broken generators form conjugate pairs, the results (\ref{the ;ast one}) and (\ref{disp}) are satisfied and in addition the result (\ref{eq:strange}) is still valid. By using then the previous arguments, we have that $M_n=M_{n'}$ ($\tilde{M}_n=\tilde{M}_{n'}$). In addition, the spatial integrations (omitted for convenience) guarantee that $\vec{p}_n=\vec{p}_{n'}$ ($\tilde{\vec{p}}_n=\tilde{\vec{p}}_{n'}$) as before. Here however, since the result (\ref{disp}) is also satisfied, then we have the conditions $M_n= \tilde{M}_n$ or $M_{n'}=\tilde{M}_{n'}$, which means that the dispersion relations for the Nambu-Goldstone bosons corresponding to conjugate pairs is the same and they represent a single degree of freedom. Taking into account the previous results, we can simplify (\ref{eq:strange}) as follows  
 
\begin{eqnarray}   \label{eq:strange3}
\sum_{n,n'}<0\vert\phi_i(z)\vert n><n\vert\ j_{0\alpha}(0)\vert n'><n'\vert j_{0\beta}(0)\vert0>e^{-i\tilde{M}_{n'}y^0}e^{i(\tilde{\vec{p}}_{n'})\cdot\vec{y}}\nonumber\\
-<0\vert\phi_i(z)\vert n><n\vert\ j_{0\beta}(0)\vert n'><n'\vert j_{0\alpha}(0)\vert0>e^{-iM_{n'}x^0}e^{i(\vec{p}_{n'})\cdot\vec{x}}\nonumber\\
-<0\vert\ j_{0\alpha}(0)\vert n'><n'\vert j_{0\beta}(0)\vert n><n\vert\phi_i(z)\vert 0>e^{iM_{n'}x^0}e^{-i(\vec{p}_{n'})\cdot\vec{x}}
\nonumber\\
+<0\vert\ j_{0\beta}(0)\vert n'><n'\vert j_{0\alpha}(0)\vert n><n\vert\phi_i(z)\vert 0>e^{i\tilde{M}_{n'}y^0}e^{-i(\tilde{\vec{p}}_{n'})\cdot\vec{y}}\nonumber\\
\neq 0.
\end{eqnarray}
Note that in this case, the summation symbols are written explicitly. Although the spatial integrations at this point are not fully developed, we have omitted the terms where such integrations will give the conditions $\vec{p}_n=\vec{p}_{n'}$, as well as $\tilde{p}_n=\tilde{p}_{n'}$. In order to simplify the notation, we can express eq. (\ref{eq:strange3}) as 

\begin{eqnarray}   \label{eq:strange4}
\sum_{n,n'}C_{n,n'}<0\vert\phi_i(z)\vert n><n'\vert j_{0\beta}(0)\vert0>e^{-i\tilde{M}_{n'}y^0}e^{i(\tilde{\vec{p}}_{n'})\cdot\vec{y}}\nonumber\\
-D_{n,n'}<0\vert\phi_i(z)\vert n><n'\vert j_{0\alpha}(0)\vert0>e^{-iM_{n'}x^0}e^{i(\vec{p}_{n'})\cdot\vec{x}}\nonumber\\
-D^*_{n,n'}<0\vert\ j_{0\alpha}(0)\vert n'><n\vert\phi_i(z)\vert 0>e^{iM_{n'}x^0}e^{-i(\vec{p}_{n'})\cdot\vec{x}}
\nonumber\\
+C^*_{n,n'}<0\vert\ j_{0\beta}(0)\vert n'><n\vert\phi_i(z)\vert 0>e^{i\tilde{M}_{n'}y^0}e^{-i(\tilde{\vec{p}}_{n'})\cdot\vec{y}}\neq 0.
\end{eqnarray}
Here once again we have defined $C_{n,n'}=<n\vert\ j_{0\alpha}(0)\vert n'>$ and $D_{n,n'}=<n'\vert j_{0\alpha}(0)\vert n>$ with the corresponding adjoint results. Here in addition $C^*_{n,n'}=C_{n',n}$ and $D^*_{n, n'}=D_{n',n}$. By taking into account that the time-(spatial) dependence of the previous expressions is translated to the exponential terms, we can factorize then the previous result as we have done for the previous case as

\begin{eqnarray}   \label{eq:strange7}
\sum_{n,n'}C_{n,n'}<0\vert\phi_i(z)\vert n><n'\vert j_{0\beta}(0)\vert0>\left(e^{-i\tilde{M}_{n'}y^0}e^{i(\tilde{\vec{p}}_{n'})\cdot\vec{y}}+e^{i\tilde{M}_{n'}y^0_R}e^{-i(\tilde{\vec{p}}_{n'})\cdot\vec{y}}\right)\nonumber\\
-D_{n,n'}<0\vert\phi_i(z)\vert n><n'\vert j_{0\alpha}(0)\vert0>\left(e^{-iM_{n'}x^0}e^{i(\vec{p}_{n'})\cdot\vec{x}}+e^{iM_{n'}x^0_R}e^{-i(\vec{p}_{n'})\cdot\vec{x}}\right).
\end{eqnarray} 
For a system which is not invariant under time-reversal, we have to distinguish between {\it forward} in time and {\it backward} in time. Then the time-reversal variables become $x^0_R=-x^0$ and $y^0_R=-y^0$ for this case. If we take into account this time-reversal condition, then we get

\begin{eqnarray}   \label{eq:strange8}
\sum_{n,n'}C_{n,n'}<0\vert\phi_i(z)\vert n><n'\vert j_{0\beta}(0)\vert0>e^{-i\tilde{M}_{n'}y^0}\left(e^{i(\tilde{\vec{p}}_{n'})\cdot\vec{y}}+e^{-i(\tilde{\vec{p}}_{n'})\cdot\vec{y}}\right)\nonumber\\
-D_{n,n'}<0\vert\phi_i(z)\vert n><n'\vert j_{0\alpha}(0)\vert0>e^{-iM_{n'}x^0}\left(e^{i(\vec{p}_{n'})\cdot\vec{x}}+e^{-i(\vec{p}_{n'})\cdot\vec{x}}\right).
\end{eqnarray} 
If we focus in the neighborhood where $\vec{p}_n\to0$, we will find that the lowest order term on the expansion is quadratic. This can be observed when we combine the terms of the form  

\begin{equation}   
e^{i(\tilde{\vec{p}}_{n'})\cdot\vec{y}}+e^{-i(\tilde{\vec{p}}_{n'})\cdot\vec{y}}=2cos(\tilde{\vec{p}}_{n'}\cdot\vec{y})\approx 2-(\tilde{\vec{p}}_{n'}\cdot\vec{y})^2+...,
\end{equation}
inside eq. (\ref{eq:strange8}). Here the quadratic dependence is evident. Note that still the frequency represented by the temporal phase goes to zero linearly. Then we have proved that in systems where pair of broken generators form conjugate pairs, the dispersion relation for the corresponding Nambu-Goldstone boson is quadratic, namely, $\omega\backsim \vec{k}^2$. We can then simplify eq. (\ref{eq:strange8}) as

\begin{eqnarray}   \label{eq:strange808}
\sum_{n,n'}C_{n,n'}<0\vert\phi_i(z)\vert n><n'\vert j_{0\beta}(0)\vert0>2cos\left(\tilde{\vec{p}}_{n'}\cdot\vec{y}\right)e^{-i\tilde{M}_{n'}y^0}\nonumber\\
-D_{n,n'}<0\vert\phi_i(z)\vert n><n'\vert j_{0\alpha}(0)\vert0>2cos\left(\vec{p}_{n'}\cdot\vec{x}\right)e^{-iM_{n'}x^0}\neq0,
\end{eqnarray} 
where the quadratic dispersion relation is more evident. Both line of terms in eq. (\ref{eq:strange808}), represent the same degree of freedom. If $C_{n, n'}=-D_{n, n'}$, then we obtain 

\begin{eqnarray}   \label{eq:strange808lala}
\sum_{n,n'}2C_{n,n'}<0\vert\phi_i(z)\vert n><n'\vert j_{0\beta}(0)\vert0>2cos\left(\tilde{\vec{p}}_{n'}\cdot\vec{y}\right)e^{-i\tilde{M}_{n'}y^0}\neq0.
\end{eqnarray} 
Note that this extra factorization is possible because of the condition $M_{n'}=\tilde{M}_{n'}$ obtained in eq. (\ref{disp}) as a consequence of the fact that we have pair of broken generators canonically conjugate. Due to this equality, we can assure that in addition $\vec{p}_{n'}=\tilde{\vec{p}}_{n'}$.  

\section{Conclusions}   \label{Sec6}

In this paper, we have obtained in detail the dispersion relation for the Nambu-Goldstone bosons for two cases. In the first case, all the broken generators are independent and then they do not form conjugate pairs. In this case, the number of broken generators is equivalent to the number of Nambu-Goldstone bosons. With the additional assumption of invariance under time-reversal, the dispersion relation for the Nambu-Goldstone bosons is linear. On the other hand, for the second case, where the Nambu-Goldstone bosons form conjugate pairs, then this pairs of canonical conjugate broken generators, represent a single degree of freedom. By writing appropriately the behavior of the system under time-reversal, then the dispersion relation is quadratic.  \\\\

{\bf Acknowledgement}
I thank Hitoshi Murayama for useful discussions about this problem during my visit to the University of California Berkeley and during the Berkeley week at the IPMU. I thank Haruki Watanabe for useful discussions during my visit to Tokyo University, Hongo campus. I thank Tomas Brauner for useful correspondences. I thank Masahito Yamazaki as well as Rene Meyer for useful discussions about this problem during my visit to IPMU. I thank to the IPMU community for the hospitality during my periodic visits to that institution, where a peaceful, open and harmonious environment can be perceived. I thank Juan Maldacena for useful discussions about this issue in the conference Strings 2016, organized in Tsinghua University. I thank the organizers of the conference Strings 2016 for giving me the chance of doing a presentation about these results inside this conference. I. A. is supported by the JSPS Post Doctoral fellow for oversea researchers.

\end{document}